\documentclass[aps,prl,superscriptaddress,showpacs,showkeys,twocolumn]{revtex4-1}

\usepackage{amsmath, amsfonts, amssymb, amsbsy}
\usepackage{graphicx}
\usepackage{xspace}
\usepackage{hyperref}
\usepackage{color}
\newcommand{\basl}{Ba$_{2}$IrO$_4$\xspace}
\newcommand{\srsl}{Sr$_{2}$IrO$_4$\xspace}
\newcommand{\bfpsi}{\mbox{\boldmath$\psi$}}

\begin{document}

\title{Robustness of basal-plane antiferromagnetic order and the $J_{\mathrm{eff}}=1/2$ state in single-layer iridate spin-orbit Mott insulators}

\author{S.~Boseggia}
\email[]{stefano.boseggia@diamond.ac.uk}

\affiliation{London Centre for Nanotechnology and Department of Physics and Astronomy, University College London, London WC1E 6BT, UK}
\affiliation{Diamond Light Source Ltd, Diamond House, Harwell Science and Innovation Campus, Didcot, Oxfordshire OX11 0DE, UK}

\author{R.~Springell}
\affiliation{Royal Commission for the Exhibition of 1851 Research Fellow, Interface Analysis Centre, University of Bristol BS2 8BS,UK}
\author{H.~C. Walker}
\affiliation{Deutsches Elektronen-Synchrotron DESY, 22607 Hamburg, Germany}
\author{H.~M. R\o{}nnow}
\affiliation{Laboratory for Quantum Magnetism, ICMP, \'Ecole Polytechnique F\'ed\'erale de Lausanne (EPFL), CH-1015 Lausanne, Switzerland.}
\author{Ch. R\"{u}egg}
\affiliation{Laboratory for Neutron Scattering, Paul Scherrer Institut, CH-5232 Villigen PSI, Switzerland.}
\affiliation{DPMC-MaNEP, University of Geneva, CH-1211 Geneva, Switzerland}
\author{H.~Okabe}
\affiliation{National Institute for Materials Science (NIMS), 1-1 Namiki, Tsukuba, Ibaraki 305-0044, Japan}
\author{M.~Isobe}
\affiliation{National Institute for Materials Science (NIMS), 1-1 Namiki, Tsukuba, Ibaraki 305-0044, Japan}
\author{R.~S. Perry}
\affiliation{Scottish Universities Physics Alliance, School of Physics, University of Edinburgh, Mayfield Road, Edinburgh EH9 3JZ, Scotland}
\author{S.~P. Collins}
\affiliation{Diamond Light Source Ltd, Diamond House, Harwell Science and Innovation Campus, Didcot, Oxfordshire OX11 0DE, UK}
\author{D.~F. McMorrow}
\affiliation{London Centre for Nanotechnology and Department of Physics and Astronomy, University College London, London WC1E 6BT, UK}

\date{\today}

\begin{abstract}
The magnetic structure and electronic groundstate of the layered perovskite \basl have been investigated using x-ray resonant magnetic scattering (XRMS). Our results are compared with those for \srsl, for which we provide supplementary data on its magnetic structure. 
We find that the dominant, long-range antiferromagnetic order is remarkably similar in the two compounds, and that the electronic groundstate in \basl, deduced from an investigation of the XRMS $L_3/L_2$ intensity ratio, is consistent with  a $J_{\mathrm{eff}}=1/2$ description.
The robustness of these two key electronic properties to the considerable structural differences between the Ba and Sr analogues is discussed in terms of the enhanced role of the spin-orbit interaction in 5$d$ transition metal oxides.
\end{abstract}



\maketitle

Transition metal oxides (TMO) containing a 5$d$ element are increasingly attracting attention as an arena in which to search for  novel electronic states\cite{Pesin,axion_PhysRevB.83.205101,PhysRevLett.105.027204,Weil_PhysRevB.85.045124}. These are proposed to derive from the strong spin-orbit interaction (SOI) in the 5$d$'s, which in essence entangles spin and orbital moments, strongly mixing spin and spatial coordinates. Iridium based compounds have featured predominantly in this quest, with considerable focus on the layered perovskites of which \srsl is the prototypical example\cite{Kim-PhysRevLett-2008}. In this case, the SOI leads to a $J_{\mathrm{eff}}=1/2$ groundstate for the Ir$^{4+}$ (5$d^5$) ions, from which a Mott-like insulator then emerges through the action of relatively weak electronic correlations which would otherwise lead to a metallic state. Direct evidence for the existence of a $J_{\mathrm{eff}}=1/2$ groundstate in \srsl was provided by x-ray resonant magnetic scatting (XRMS) experiments which revealed a much stronger resonance at the $L_3$ edge than at the $L_2$\cite{Kim-Science-2009}.

\begin{figure}
\centering
\includegraphics[width=0.5\textwidth]{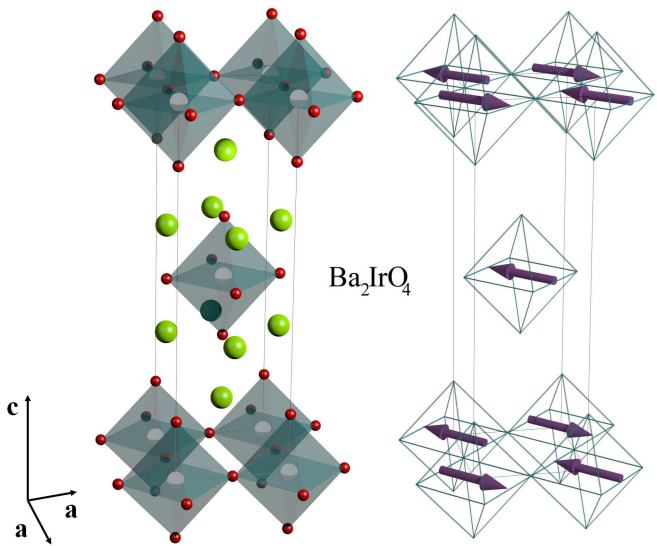}
\caption{ (Color Online) The left-hand panel shows the crystal structure of \basl. Perovskite IrO$_6$ layers, where the Ir atoms (grey) lie at the center of corner sharing oxygen (red) octahedra, are separated by Ba atoms (light green). The right-hand panel shows the basal-plane antiferromagnetic structure of the \basl where the magnetic moments are pointing along the [1\,1\,0] direction.}
\label{fig:MagStruct}
\end{figure}

The structural similarity of the single-layer iridates to La$_2$CuO$_4$ adds further impetus to the study of these materials, opening as it does a possible route to the discovery of new families of superconductors\cite{PhysRevLett.106.136402}.
In this context, a particularly interesting compound is \basl, since  structurally it is a closer 5$d$ analogue of  La$_2$CuO$_4$ than the Sr compound.
\basl crystallizes in the K$_2$NiF$_4$-type structure (space group I4/\textit{mmm}) with 180$^\circ$ Ir-O-Ir bonds
in the basal-plane (Fig.\ \ref{fig:MagStruct}), and with a 7\% tetragonal distortion of the IrO$_6$ octahedra along the [0\,0\,1] direction\cite{PhysRevB.83.155118}. In contrast, in \srsl (I4$_1$/\textit{acd}), there is a staggered, correlated rotation of the IrO$_6$ octahedra by 11$^\circ$, and a tetragonal distortion of 4\%\cite{PhysRevB.49.9198}.

From a theoretical point of view, both the tetragonal distortion and the presence or otherwise of octahedral rotations have significant consequences for the electronic and magnetic properties. Firstly, it should be noted that the $J_{\mathrm{eff}}=1/2$ state itself is only strictly realized in a system of cubic symmetry\cite{abragam1970electron}. Secondly, the loss of inversion symmetry in \srsl gives rise to a finite Dzyaloshinskii-Moriya (DM) interaction, allowing the formation of non-collinear magnetic structures\cite{PhysRevLett.102.017205}. Both of these effects on the magnetism in \basl and \srsl have been investigated using \textit{ab-initio} methods\cite{PhysRevB.85.220402}.

For \srsl, the consequences of these structural features for the electronic and magnetic properties have been comprehensively explored in a number of experimental and theoretical studies\cite{Kim-Science-2009,Kim-PhysRevLett-2008,PhysRevLett.102.017205,PhysRevLett.108.247212}.
By contrast, for \basl  there are a number of important open questions, including whether or not its groundstate can reasonably be assigned as  $J_{\mathrm{eff}}=1/2$, and
the exact nature of its magnetic structure. The latter question is of particular relevance to the prospect of \basl becoming the parent compound of a new family of unconventional, magnetically mediated superconductors. Both cuprate and pnictide superconductors, for example, emerge when doping destabilizes long-range antiferromagnetic order, and in each case obtaining a microscopic understanding
of the magnetic groundstate of the parent compound has played a pivotal role in our knowledge\cite{RevModPhys.78.17,Zhao2008}. From a range of bulk probes and muon spin rotation ($\mathcal{\mu}$SR) it is known that \basl exhibits a magnetic transition below $\sim$\,240~K\cite{PhysRevB.83.155118}, close to the magnetic transition in \srsl of $T_\mathrm{N}$\,$\sim$\,230~K, below which the magnetic moments in \srsl form  a canted antiferromagnetic structure\cite{Kim-Science-2009}. Whether or not the ferromagnetic moment resulting from the canting is inimical for superconductivity when \srsl is doped to form a metal is another important open question.

In this letter we report the results of our XRMS investigation of \basl, which addresses both the question of the magnetic structure in \basl, and the relevance of the $J_{\mathrm{eff}}=1/2$ description to its electronic groundstate. Our results are compared with corresponding measurements on \srsl, for which we also supply supplementary data, and discussed in terms of current theoretical models. The major
achievement of our study is to establish that both antiferromagnetic order and the $J_{\mathrm{eff}}=1/2$ state are, to a remarkable degree, robust to structural distortions in the single layered iridate perovskites.

Single crystals of \basl were synthesized at the National Institute for Materials Science (NIMS) by the slow-cooling technique under pressure. The sample of size $\sim$ 200$\mu$m $\times$ 200 $\mu$m $\times$ 200 $\mu$m,  was initially checked with a Supernova x-ray diffractometer using a monochromatic Mo source at the Research Complex at Harwell (RCaH), Chilton, UK. The diffraction data are consistent with the I4/\textit{mmm} space group and cell parameters a\,=\,b\,=\,4.0223(4) \AA\ and c\,=\,13.301(3) \AA\ at room temperature. The \srsl single crystals were prepared at the University of Edinburgh following the standard self-flux technique\cite{PhysRevB.57.R11039}. The correlated rotation of the IrO$_6$ octahedra about the $c$ axis reduces the space group symmetry to I4$_1$/\textit{acd}, generating a larger unit cell: $\sqrt{2}a\times \sqrt{2}b\times 2c$, under the rotation of the original cell by 45$^\circ$\cite{PhysRevB.49.9198}.   The XRMS measurements were performed at the Ir $L_{2}$ (12.831 keV) and $L_{3}$ (11.217 keV) edges, probing dipolar transitions from $2p_{\frac{1}{2}}$ to $5d$ and from $2p_{\frac{3}{2}}$ to $5d$, respectively. The experiment on the \basl crystal was conducted at the I16 beamline of the Diamond Light Source, Didcot, UK.
X-rays were focussed to a  beam size of 20$\times$200 $\mu$m(V$\times$H) at the sample position. The sample was mounted in a Displex cryostat with the [1\,1\,0] and [0\,0\,1] directions in the vertical scattering plane. In order to discriminate between different scattering mechanisms, an Au (3\,3\,3) polarization analyzer was exploited for the entire energy range (11.217 keV-12.831 keV). The XRMS study on  \srsl was performed at the P09 beamline\cite{P09} of Petra III, at DESY, Germany.
On P09 the x-rays were focused to a beam size of $50\times50$ $\mu$m at the sample position, using a set of focusing mirrors and beryllium compound refractive lenses. The sample was mounted in a Displex cryostat with the [1\,0\,0] and [0\,0\,1] directions in the vertical scattering plane. A pyrolytic graphite (0\,0\,8) crystal was exploited to analyze the polarization of the scattered beam.

\begin{figure}
\centering
\includegraphics[width=1\columnwidth]{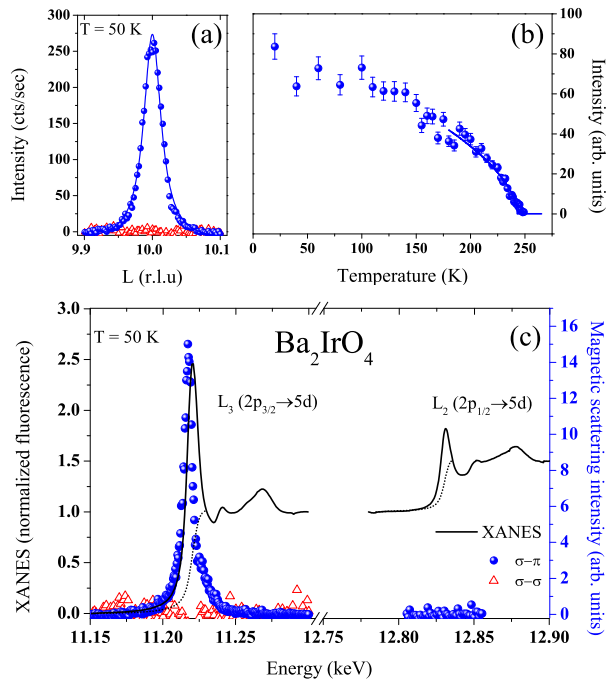}
\caption{ (Color online) (a) $L$ scans across the ($\frac{1}{2}$\,$\frac{1}{2}$\,$10$) magnetic reflection at the Ir $L_3$ edge, $T$ = 50 K in \basl. (b) The temperature dependence of the ($\frac{1}{2}$\,$\frac{1}{2}$\,10) magnetic reflection at the Ir $L_3$ edge in \basl. The solid blue line is a fit to a power law. (c) Resonant enhancement of the ($\frac{1}{2}$\,$\frac{1}{2}$\,$8$) magnetic reflection across the $L_{2,3}$ edges at $T$ = 50 K in \basl. The solid black line shows the x-ray absorption near edge structure (XANES) spectra, measured in fluorescence mode, normalized to the number of initial states. The blue spheres and red triangles show the intensity of the ($\frac{1}{2}$\,$\frac{1}{2}$\,$8$) reflection. The black dashed line demarcates the integrated white line used to calculate the branching ratio.}
\label{fig:En}
\end{figure}

In \basl, with the photon energy tuned to be close to the $L_3$ edge (11.222 keV) and the sample cooled to 50~K, sharp peaks were found at the reciprocal lattice points
($\frac{1}{2}$\,$\frac{1}{2}$\,$L$) with $L$ even. These peaks existed in the rotated photon polarization channel $\sigma-\pi$ only (see Fig.~\ref{fig:En}(a)) as expected from the selection rules for XRMS arising from dipolar transitions\cite{Hill:sp0084}. We thus deduce that the Ir$^{4+}$
magnetic moments order in an antiferromagnetic structure, with a doubling of the unit cell along the in-plane directions, described by a magnetic propagation vector of $\mathbf{k}$\,=\,[$\frac{1}{2}$\,$\frac{1}{2}$\,$0$].

In Fig.~\ref{fig:En}(c) we present the energy dependence of the magnetic scattering at ($\frac{1}{2}$\,$\frac{1}{2}$\,$8$) together with x-ray absorption near edge structure (XANES) measurements for energies in the vicinity of the $L_3$ and $L_2$ edges. The most notable features of this data are the existence of a well-defined resonance at the $L_3$  edge, and the complete absence of a response at the $L_2$ edge within experimental uncertainty. Concerted attempts to find a magnetic response at the $L_2$ edge by investigating various magnetic reflections all ended in failure. In their study of \srsl, \textcite{Kim-Science-2009} argued that the observed large XRMS intensity ratio, I$_{L_3}$/I$_{L_2}$, served as a unique fingerprint of the $J_{\mathrm{eff}}=1/2$ state, since for the pure $J_{\mathrm{eff}}=1/2$ state I$_{L_2}$ is identically zero. Our results, interpreted in this spirit,
establish that even in the presence of a large tetragonal distortion, \basl belongs to the same class of $J_{\mathrm{eff}}=1/2$ spin-orbit Mott insulators as \srsl.

The width of the $L_3$ resonance is FWHM$_{L_{3}}$\,=\,7.6(1) eV, comparable to the values found in Sr$_2$IrO$_4$ and in Sr$_3$Ir$_2$O$_7$\cite{Kim-Science-2009,PhysRevB.85.184432}. The position of the resonance, similarly to those of \srsl and Sr$_3$Ir$_2$O$_7$, is 3 eV below the $L_3$ white line. From the analysis of the XANES spectra we find a very large branching ratio BR = 5.45\footnote{For a more detailed analysis of the branching ratio in \basl see Supplemental Material.}. This is a further confirmation of the strong SOI regime in \basl.

The thermal evolution of the antiferromagnetic order was determined by performing $\theta-2\theta$ scans of the ($\frac{1}{2}$\,$\frac{1}{2}$\,10) reflection in the $\sigma-\pi$ channel  at the energy (11.219 keV) that maximizes the XRMS response. Fig.~\ref{fig:En}(b) shows the integrated intensity obtained  by fitting a Lorentzian peak shape to the individual scans as a function of temperature. The transition appears to be second order, and from the fit to a  $A(1-\frac{T}{T_\mathrm{N}})^{2\beta}$ function we obtain the Neel temperature $T_\mathrm{N}$\,=\,243(1) K, in excellent agreement with the value found by $\mathcal{\mu}$SR measurements\cite{PhysRevB.83.155118}.

In order to determine  the possible magnetic structures in \basl, we performed representation analysis by means of the SARA\textit{h}\cite{AS2000680} package.
The input parameters were the system space group I4/\textit{mmm}, the magnetic propagation vector $\mathbf{k}$\,=\,[$\frac{1}{2}$\,$\frac{1}{2}$\,$0$], resulting from the XRMS measurements, and the atomic coordinates of the Ir atoms.
The results of the SARA\textit{h} calculations are presented in Table \ref{basis_vector_table_1}. For Ba$_2$IrO$_4$ only 3 irreducible representations (IR's), with the associated basis vectors, are possible: $\Gamma_3$, $\Gamma_5$ and $\Gamma_7$. Contrary to \srsl, the symmetry of the system, that preserves the inversion symmetry, rules out any representation that involves a ferromagnetic component.
\begin{table}
\begin{ruledtabular}
\begin{tabular}{ccc|cccccc}

  IR  &  BV  &  Atom & \multicolumn{6}{c}{BV components}\\
      &      &             &$m_{\|a}$ & $m_{\|b}$ & $m_{\|c}$ &$im_{\|a}$ & $im_{\|b}$ & $im_{\|c}$ \\
\hline
$\Gamma_{3}$ & $\bfpsi_{1}$ &      1 &      0 &      0 &      1 &      0 &      0 &      0  \\
$\Gamma_{5}$ & $\bfpsi_{2}$ &      1 &      1 &      1 &      0 &      0 &      0 &      0  \\
$\Gamma_{7}$ & $\bfpsi_{3}$ &      1 &      1 &     -1 &      0 &      0 &      0 &      0  \\
\end{tabular}
\end{ruledtabular}
\caption{Basis vectors for the space group I4/\textit{mmm} with
${\bf k}$ = [$\frac{1}{2}$\,$\frac{1}{2}$\,$0$].The
decomposition of the magnetic representation for
the Ir site
$( 0,~ 0,~ 0)$ is
$\Gamma_{Mag} = 0\Gamma_{1}^{1}+0\Gamma_{2}^{1}+1\Gamma_{3}^{1}+0\Gamma_{4}^{1}+1\Gamma_{5}^{1}+0\Gamma_{6}^{1}+1\Gamma_{7}^{1}+0\Gamma_{8}^{1}$. The atom of the primitive
basis is defined according to
1: $( 0,~ 0,~ 0)$.}
\label{basis_vector_table_1}
\end{table}
\begin{figure}[h!]
\centering
\includegraphics[width=1\columnwidth]{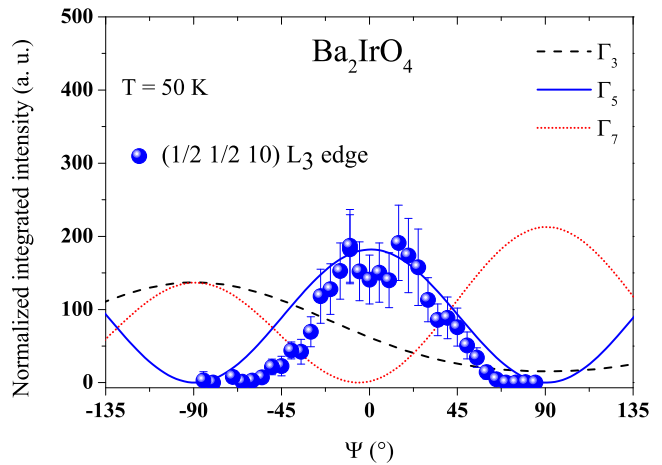}
\caption{ (Color Online) The azimuthal dependence of the ($\frac{1}{2}$\,$\frac{1}{2}$\,10) magnetic reflection (solid blue spheres) at the Ir $L_3$ edge, $T$ = 50 K in \basl . The solid lines are the azimuthal dependencies calculated for the three different IR's. The azimuthal angle $\Psi$ is defined with respect to the reference vector [1\,1\,0] in the I4/\textit{mmm} space group.}
\label{fig:Azi}
\end{figure}

To discriminate between the 3 possible structures, we performed azimuthal scans of the ($\frac{1}{2}$\,$\frac{1}{2}$\,10) magnetic reflection at the Ir $L_3$ edge, $T$\,=\,50 K. This method consists in measuring $\theta-2\theta$ scans for different $\Psi$ angles, rotating the sample around the scattering vector. From the azimuthal modulation of the intensity of the XRMS signal it is possible to determine the orientation of the magnetic moments in an antiferromagnetic material\cite{0953-8984-24-31-312202}.
Fig.\ref{fig:Azi} shows the azimuthal dependence of the ($\frac{1}{2}$\,$\frac{1}{2}$\,10) reflection (blue solid points). The dashed black line, solid blue line and dotted red line are the  azimuthal dependence for the  $\Gamma_3$, $\Gamma_5$ and $\Gamma_7$ IR, respectively, calculated by means of the FDMNES package\cite{PhysRevB.63.125120}. The experimental curve most closely resembles the calculation for the $\Gamma_5$ representation. We therefore conclude unambiguously that \basl exhibits a basal-plane antiferromagnetic order with the magnetic moments pointing along the [1\,1\,0] direction. The magnetic structure of \basl is shown in Fig.~\ref{fig:MagStruct}.

\begin{figure}
\centering
\includegraphics[width=1\columnwidth]{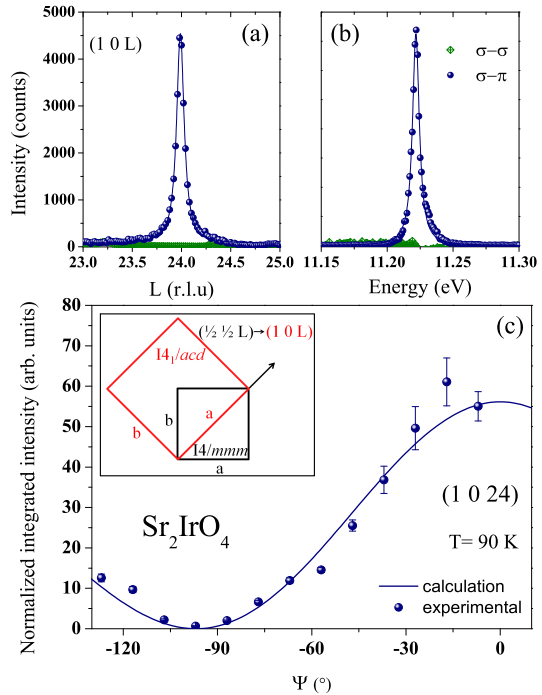}
\caption{ (Color Online) Reciprocal space $L$-scan (a) and energy dependence (b) of the XRMS intensity  of the (1 0 24) reflection in \srsl at the Ir $L_3$ edge, $T$ = 90 K. The solid blue line is a fit to a Lorentzian peak shape. The azimuthal dependence of the same reflection (c) is compared with a calculation for collinear moments along [1\,0\,0]. The azimuthal angle $\Psi$ is defined with respect to the reference vector [1\,0\,0] in I4$_1$/\textit{acd}, which corresponds to the [1\,1\,0] in the I4/\textit{mmm} space group, as demonstrated in the inset by means of a 2D projection onto the basal-plane of the unit cell.}
\label{fig:Sr}
\end{figure}

To understand the dependence of the $J_{\mathrm{eff}}=1/2$ state and the associated Hamiltonian on symmetry and lattice distortions, we have investigated the magnetic structure of \srsl. In particular we focus on the polarization and azimuthal dependencies of the XRMS, neither of which have been reported\cite{Kim-Science-2009}. With the photon energy tuned to the Ir $L_3$ edge, well defined magnetic peaks were found at the (1\,0\,$4n$) and (0\,1\,4$n$+2) Bragg positions, which are forbidden within the I4$_1$/\textit{acd} space group and correspond to the ($\frac{1}{2}$\,$\frac{1}{2}$\,$L$) peaks observed in the \basl (as illustrated in the inset of Fig.~\ref{fig:Sr}(c)). Fig.~\ref{fig:Sr}(a-b) shows the $L$ scan and the energy scan of the (1\,0\,24) magnetic reflection at the Ir $L_3$ edge at $T$\,=\,90 K. The well defined $L$ scan supports the existence of a long-ranged antiferromagnetic order. The Lorentzian shape of the energy scan (FWHM$_{L_3}$\,=\,6.26(9) eV) and the absence of any $\sigma -\sigma$ scattering mechanism confirms the magnetic nature of the peaks, similarly to \basl. These results are in agreement with the first XRMS study of \srsl\cite{Kim-Science-2009}.

In order to determine the direction of the magnetic moments in \srsl we performed azimuthal scans at the Ir $L_3$ edge, $T$ = 90 K.
The results, together with the FDMNES calculation using the same moment direction as in the irreducible representation $\Gamma_5$ of \basl, are presented in Fig.~\ref{fig:Sr}(c). Note the equivalence of the $\Psi$ angles in Fig.~\ref{fig:Sr} with those in Fig.~\ref{fig:Azi}, for the correspondence of the [1\,1\,0] direction in I4/\textit{mmm} to the [1\,0\,0] direction in the I4$_1$/\textit{acd}. By comparing the azimuthal dependence of the (1\,0\,24) reflection in \srsl with the azimuthal dependence of ($\frac{1}{2}$\,$\frac{1}{2}$\,10) reflection in \basl, we deduce that in \srsl the antiferromagnetic component is oriented along the [1\,1\,0] direction of the I4/\textit{mmm} reference system. We therefore conclude that the two compounds have essentially the same basal-plane antiferromagnetic
structure\footnote{We note that XRMS does not couple to a canted component since the latter can be seen as a ferromagnetic modulation of the antiferromagnetic structure. As a consequence, this scattering mechanism occurs in the same position in the reciprocal lattice as the charge scattering.}.

We now turn to the discussion of our results.
An effective, low-energy Hamiltonian for the layered iridates, valid in the strong SOI limit, incorporating both the effects of a tetragonal  crystal field and rotation of the IrO$_6$
octahedra (by an angle $\alpha$), has been derived by \textcite{PhysRevLett.102.017205}, which we write as
\begin{equation}
\label{H}
\mathcal{H}_{ij}=J\overrightarrow{S_i}\cdot\overrightarrow{S_j}+ J_z S_i^zS_j^z+D\cdot\left[\overrightarrow{S_i}\times\overrightarrow{S_j} \right]+\mathcal{H}^\prime.
\end{equation}
The terms on the righthand side are the isotropic Heisenberg exchange, the symmetric and asymmetric DM anisotropies, and finally an anisotropic contribution from the
Hund's coupling\cite{PhysRevLett.102.017205}. This Hamiltonian has been used to successfully account for the canted magnetic structure observed in \srsl\cite{Kim-Science-2009}, and additionally for a
dimensionally driven spin reorientation in its bi-layer counterpart Sr$_3$Ir$_2$O$_7$\cite{PhysRevLett.109.037204,PhysRevB.85.184432,0953-8984-24-31-312202}. For \basl, it would also seem to offer a natural explanation of our results: with $\alpha$\,=\,0, the second
and third terms are identically zero, leaving a leading isotropic exchange along with a weaker anisotropy, a Hamiltonian that readily supports the commensurate antiferromagnetic
order observed in our experiments. One important proviso, however, is that the magnetic groundstate supported by this Hamiltonian becomes unstable above a critical value of tetragonal distortion leading to a spin reorientation where the moments point along the [0\,0\,1] direction. Nevertheless it  seems, that nearly doubling the tetragonal distortion in moving from Sr to Ba is insufficient to exceed the critical threshold.

Although the above analysis provides a general framework for us to understand the formation of magnetic  structures in the layered perovskites,
and most especially the canting of the moments in \srsl, it does not address the key fact revealed in our experiments that the antiferromagnetic
components in the two compounds are essentially identical. For this we refer to explicit calculations of $J$ by \textcite{PhysRevB.85.220402}, who
used an \textit{ab-initio} many-body approach.
Their calculations show that when the SOC is switched off, the groundstate and the magnetic interactions are extremely sensitive to the local symmetry and so very different in the two systems: \basl has a hole in the $d_{\mathrm{xz}}/d_{\mathrm{yz}}$ states and a strong antiferromagnetic $J$ interaction ($\sim$~~15.4 meV), \srsl has a hole in the $d_{\mathrm{xy}}$ state and a ferromagnetic $J$ interaction ($\sim$~--19.2 meV). However, upon including the SOC, the hole acquires an equal $d_{\mathrm{xy}}$, $d_{\mathrm{zx}}$ and $d_{\mathrm{yz}}$ character in both compounds and $J$ in \srsl becomes antiferromagnetic ($\sim$~51.3 meV), and almost identical to that in \basl ($\sim$~58 meV).
Therefore, the robustness of antiferromagnetic order in the layered perovskites to structural distortions, is ultimately linked to the strong SOI, which produces a groundstate wavefunction that is three dimensional and inherently less perturbed by structural distortions.

In this letter we have presented a detailed XRMS study of the magnetic and electronic structures  of the single layered iridates \basl and \srsl. \basl is found to be a basal-plane commensurate antiferromagnet below $T_\mathrm{N}$\,=\,243 K. Azimuthal scans combined with group theory calculations have been employed to prove that the moments order along the [1\,1\,0] direction. From a comparison with XRMS data on the related compound \srsl, we establish that both compounds have essentially the same basal-plane antiferromagnetic structure, in spite of their structural differences. We also conclude from our results for the intensity  ratio $L_3/L_2$ of the XRMS signal that \basl is also in the same class of $J_{\mathrm{eff}}=1/2$ spin-orbit Mott insulators as \srsl.  Thus both the magnetic and electronic structures in the layered perovskites are remarkably robust to structural distortions, a  fact that can be linked
directly to the unique three-dimensional character of the $J_{\mathrm{eff}}=1/2$ state produced by the strong SOI which renders it  insensitive to the perturbations in local symmetry.

We would like to thank the Impact studentship programme, awarded jointly by UCL and Diamond Light Source for funding the thesis work of S. Boseggia. G. Nisbet provided excellent instrument support and advice on multiple scattering at the I16 beamline.  We also thank J. Strempfer and D. K. Shukla for technical support at P09.  The research was supported by the EPSRC, and part of the research leading to these results has received funding from the European Community's Seventh Framework Programme (FP7/2007-2013) under grant agreement n$^{\circ}$ 312284.

\bibliography{Ba214}

\end{document}